# Imaging blood cells through scattering biological tissue using speckle scanning microscopy

Xin Yang, Ye Pu, and Demetri Psaltis[*]

*Optics Laboratory, Ecole Polytechnique Frederale de Lausanne (EPFL), Lausanne, 1015, Switzerland*
[*]*demetri.psaltis@epfl.ch*

**Abstract:** Using optical speckle scanning microscopy [1], we demonstrate that clear images of multiple cells can be obtained through biological scattering tissue, with subcellular resolution and good image quality, as long as the size of the imaging target is smaller than the scanning range of the illuminating speckle pattern.

## 1. Introduction

Bertolotti et al. [1] recently reported a method for imaging fluorescent objects behind scattering media. The technique is based on scanning the speckle pattern produced by the scatterers across the object and collecting the resulting fluorescence back through the scattering medium. We refer to this method as speckle scanning microscopy (SSM). SSM is one of the many methods recently reported to generate optical images in scattering media. These methods [2-19] expand earlier efforts on phase conjugation [20] and imaging with ballistic photons [21] using modern tools for digital detection and control of the amplitude and phase of optical fields. SSM is unique among all these methods in that it relies on the scattering medium itself to read the object rather than using a beacon or some other ways to measure the system. As a result, SSM makes it possible for an object behind a scattering screen to be imaged without any information from the same side of the screen.

Speckle scanning microscopy relies on the correlation in the speckle intensity, a phenomenon often termed as the optical memory effect [22]. In this paper, we elucidate the role of the memory effect and key parameters in the implementation of SSM using numerical simulations. We show that within these constraints SSM can be used to image blood cells enclosed in scattering biological tissues. The images we obtained were of good quality, demonstrating that SSM has a potential in biological applications.

## 2. Theoretical analysis

The principle of SSM is illustrated in Fig. 1. A laser beam of diameter $W$ is injected into the first scattering medium of thickness $L_1$, which projects a speckle intensity field $S$ on the sample plane after propagation over distance $d_1$. Rotating the beam by an angle $\theta$ causes the speckle pattern to shift by $q$ in the two-dimensional sample plane, where $q = d_1\theta$. The fluorescence excited by the speckle pattern in the sample $O$ propagates through distance $d_2$ and is scattered again by the second scattering medium of thickness $L_2$, after which the fluorescence signal is filtered and detected. The averaged autocorrelation $\langle I \otimes I \rangle(q)$ of the measured intensity map can be expressed as [1] $\langle I \otimes I \rangle(q) = \langle [(O \otimes O) * (S \otimes S)](q) \rangle$, where $*$ is the convolution operator, and $\otimes$ is the correlation operator. When scanning the incident angle, the random speckle pattern that excites the object fluorescence scans along, albeit with a lower correlation. Laser speckles have a well-defined autocorrelation profile [23], which effectively becomes a delta function when the scanning step is larger than the mean speckle size. This renders $\langle I \otimes I \rangle(q)$ approximately equal to the object autocorrelation $\langle O \otimes O \rangle(q)$, from which it is possible to recover $O$ with a phase retrieval algorithm [24]. In practice, each particular realization of the speckle pattern introduces noise in $I \otimes I(q)$, and the decreasing correlation over the increasing angle also introduces an uncorrelated speckle pattern. Hence multiple instances of the autocorrelation profile are averaged to suppress random noise but the uncorrelated portion of the speckle illumination remains as a major factor deteriorating the quality of the image.

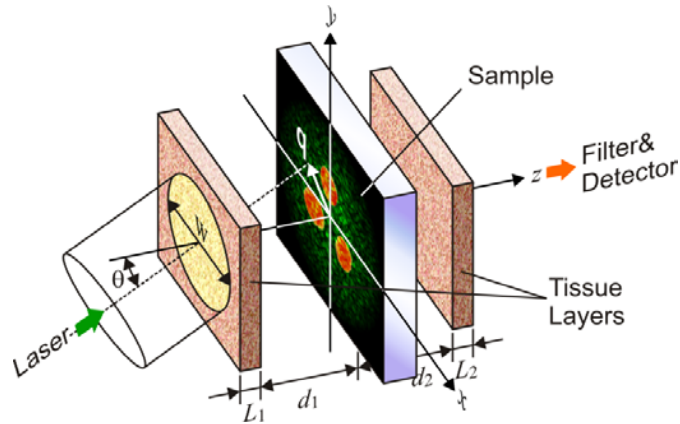

Fig. 1. Schematic diagram of SSM setup

For subwavelength scatterers, Feng et al [22] gave an analytical solution to the memory effect, which can be expressed as $C(\theta) = (k\theta L_1)^2 / \sinh^2(k\theta L_1)$ (Fig. 2 solid line). This limits the maximum incident angle to roughly $\theta_{max} = 2/kL_1$, beyond which the speckle correlation is less than 30%. In biological tissues, however, the scattering properties are more complicated, with contributions coming from both cellular level structures and subwavelength organelles. The correlation curves as a function of incident angle were measured for a series of biological tissues and plotted in Fig. 2. $\theta_{max}$ for the 50 μm tissues are in general larger than the theoretical prediction made by Feng, indicating that cellular level scatterers play an important role in the scattering properties of biological tissue.

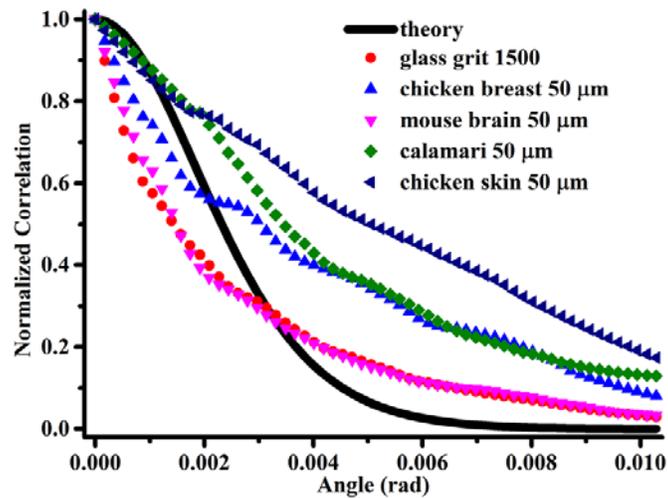

Fig. 2. Normalized correlation verses tilting angle for different tissues

The scanning range $2q_{max}$ is determined by $d_1$ and $\theta_{max}$, where $2q_{max} = 2d_1\theta_{max}$. Given the small $\theta_{max}$ (around 3-9 mrad, read from Fig. 2), a large $d_1$ is required to achieve a reasonable $2q_{max}$. On the other hand, the physical resolution of SSM, determined by the mean speckle size, is $\lambda d_1/W$ [23], in the case of $W \ll d_1$. A large $d_1$ is therefore unfavorable in terms of resolution. Based on these facts, the choice of $d_1$ is a vital factor that determines the performance of SSM.

Two conditions need to be satisfied, in order to obtain undistorted images of fluorescence objects in SSM: 1. The size of the entire fluorescent object needs to be smaller than the scanning range; 2. No other objects outside the scanning range can be excited by the speckle pattern. Condition 1 can be satisfied by controlling $d_1$, but condition 2 can be more problematic in practice due to other fluorescence sources in close proximity to the object. Such fluorescence sources would be illuminated by a speckle pattern that is uncorrelated with the original speckle pattern. Yet, the fluorescence from them contributes to the intensity map, and confuses the image recovery algorithm. The effect coming from the fluorescence sources out of scanning range was numerically investigated in Fig. 3. The parameters in the simulation were chosen according to the realistic conditions in experiments.

Fig. 3 (a) shows an image of 8 florescent beads, arranged in an area of 128 μm by 128 μm. The speckle pattern was simulated by the following three steps: 1. Generated certain number of random distributed point sources; 2. Calculated the spherical field created by each point source; 3. Added up the fields at the distance of $d_1$. The memory effect was approximated by a linear combination of two uncorrelated speckle fields with different weighting, and the weighting values were chosen so that the calculated memory effect curve fitted the actual data measured in the experiment (Fig. 3 chicken skin 50 μm). The scanning range $2q_{max}$ was 8 mrad and kept constant. $d_1$ needs to be minimum 8 mm to satisfy the conditions 1 and 2. Fig. 3 (b)-(d) show the simulation results when $d_1 = 8$ mm. Fig. 3 (b) is the calculated autocorrelation of the original object, (c) is the averaged autocorrelation of 30 scanning intensity maps, and (d) is a retrieved object from (c). Fig. 2 (d) faithfully recovers the original object.

However, the reconstruction quality deteriorates as $d_1$ decreases. Fig. 3 (e)-(g) represent the simulation results when $d_1 = 6$ mm, and the scanning range is 96 μm. In this case, the object is the part within the red square of Fig. 3 (a). Signal coming from the fluorescence beads outside the red square are collected as well, and introduce unwanted patterns in the averaged autocorrelation (Fig. 3 (f)) and reconstructed image (Fig. 3 (g)). This phenomenon becomes more prominent if $d_1$ further decreases to $d_1 = 4$ mm and $d_1 = 2$ mm, and the corresponding results are presented in Fig. 3 (h)-(j) and Fig. 3 (k)-(m).

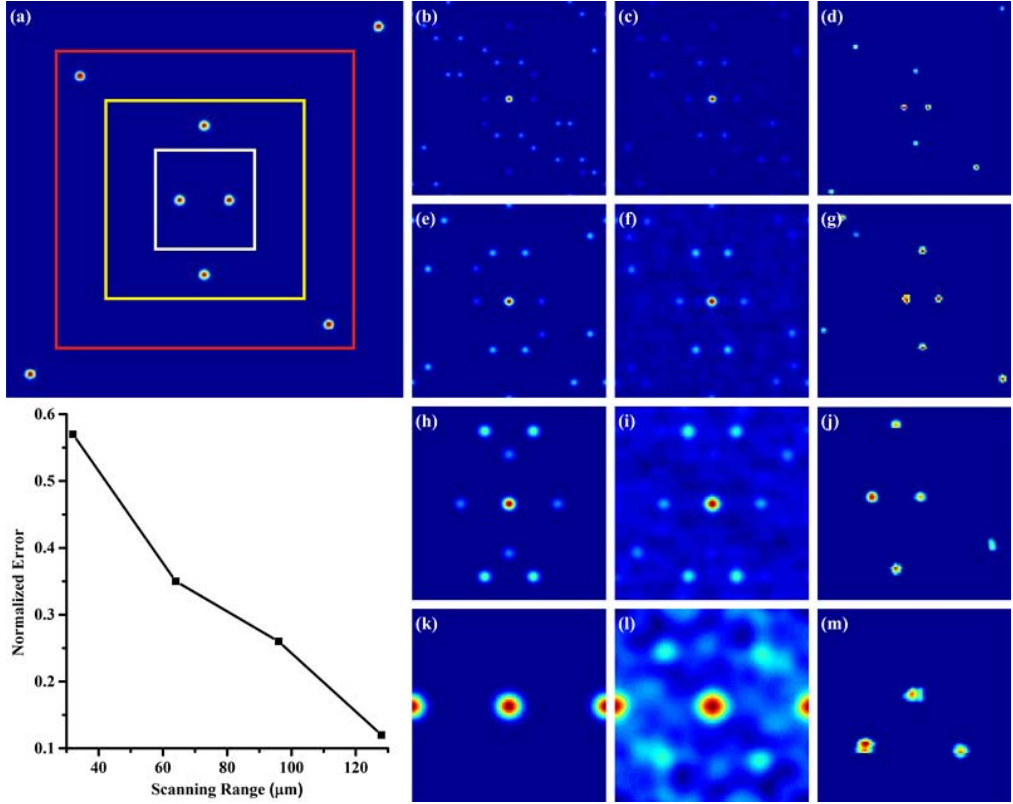

Fig. 3 (a) original imaging object (128×128 μm$^2$); (b)-(d): scanning range 128 μm, covering entire (a); (b) calculated autocorrelation of the object covered by the scanning range; (c) averaged autocorrelation of 30 scanning intensity maps; (d) a retrieved object from (c); (e)-(g): scanning range 96 μm, covering parts within red square of (a); (h)-(j): scanning range 64 μm, covering parts within yellow square of (a); (k)-(m): scanning range 32 μm, covering parts within white square of (a); (n) normalized error (between the reconstructed object and the original object covered by the scanning range) varies with scanning range

The effect from the fluorescence sources outside the scanning range can be understood by considering the speckle pattern as a combination of several overlapping uncorrelated speckle fields spaced one memory range from each other in the space of incident angle. Each of these shifted speckle patterns is weighted with the corresponding memory coefficient and produces independently an autocorrelation. Since each autocorrelation is centered, the resulting autocorrelation of the measured intensity is a superposition of the autocorrelation of multiple partial objects. Therefore, an effect akin to "aliasing" occurs. Just as in aliasing this part of the signal still contains potentially useful information about the object, but it appears as distortion and it confuses the object retrieval algorithm.

The signal-to-noise ratio (SNR) is also a crucial factor. The signal comes from the fluctuation of fluorescence intensity when the speckles scan across the object. The object fluorescence is proportional to the number of speckles falling on the object, which is roughly $A_o/A_s$, where $A_s$ is the mean speckle size and $A_o$ is the total object size. Its fluctuation relative to the mean value is then approximately $(A_s/A_o)^{1/2}$. The total number of photons $N$ received by the detector must ensure that the signal fluctuation exceeds the shot noise, i.e. $N > A_o/(\eta^2 A_s)$, where $\eta$ is the detector quantum efficiency. Therefore, smaller objects are of a clearly advantage in terms of SNR over larger objects. The detector quantum efficiency, which in a broader sense also includes the collection efficiency of the optics, has an even stronger impact to the SNR due to the second-order dependency.

## 3. Experiments and results

Having established an understanding of the conditions to obtain undistorted images in SSM, we proceeded to design an experiment to demonstrate the application of SSM in the imaging of blood cells enclosed by chicken skin tissues. A laser beam of 488 nm wavelength was delivered to the sample through a pair of galvo mirrors for two-dimensional angular scanning. The scanning system was assisted by two pairs of conjugated 1:1 relay lenses such that the beam position remained unchanged while scanning its angle. The beam size at the entrance face of the sample was 2 mm. The sample consisting of Eosin-Y stained blood cells (see methods) was placed in-between two layers of chicken skin tissue with thickness of 50 μm (Fig. 1). The distances between the blood sample to the first and second tissue layer was 4 mm and 0.5 mm, respectively. The mean speckle size projected on the blood sample was roughly 1 μm.

Given by the memory effect curve shown in Fig. 2, the maximum scanning range was roughly 64 μm. The angular scanning step was chosen as 0.12 mrad, corresponding to ~0.5 μm/step in the plane of the blood sample. The fluorescence signal, after propagating through the second layer of the scattering tissue, was collected by a microscope objective (50× 0.7 NA) and filtered with a 483 nm notch filter (20 nm bandwidth) and a long pass filter cut at 532 nm. The combined rejection at the excitation wavelength exceeded an optical density of 14. An electron-multiplying charge-coupled device (EMCCD) was used as the detector in this demonstration for ease of alignment. The intensity captured by the entire EMCCD at one incident angle was summed up as one value, and an intensity map was generated by plotting different incident angles versus their corresponding intensity values. In practical application, a single-element photo-detector such as a photo-multiplying tube (PMT) or an avalanche photodiode (APD) could be used instead of EMCCD for higher speed and lower cost.

30 scanning intensity maps were acquired and their autocorrelations were averaged. For each scan, distinct non-overlapping incident angles were used to achieve uncorrelated speckles. The average autocorrelation pattern was thresholded, and the image of the object was retrieved from the autocorrelation by applying Fienup's iterative phase retrieval algorithm. The autocorrelation was thresholded at ~5% of the autocorrelation peak, to remove the parts which change rapidly from one scan to the next, and further minimize the statistical noise. The optimum threshold was determined by trial and error and it varied from sample to sample.

Fig. 4 (a) shows the wide field fluorescence image (with scattering media removed) of a white blood cell. Fig. 4 (b) is the calculated autocorrelation of (a), from which the phase retrieval algorithm is able to recover an image of the original object (c) with excellent fidelity. Shown in Fig. 4 (d) is a typical two-dimensional fluorescence intensity map as measured with the SSM. From 30 instances of such intensity map the average autocorrelation was calculated, as displayed in Fig. 4 (e). Fig. 4 (f) shows the reconstructed image of the original object from the average autocorrelation (e). The recovered image (f) bears close resemblance to the original image, despite of some loss of fidelity due to statistics and optical noises. Nevertheless, the acuity of Fig. 4 (f) is sufficient to clearly identify the white blood cell and measure its dimension and approximate shape. The pixel resolution of Fig. 4 (d)-(f) is 101×101 pixels, which corresponds to an actual scanning range of 50 μm, smaller than the maximum scanning range of 64 μm. The object size is roughly 10-15 μm, and no other cells or fluorescence sources exists in its close proximity.

Fig. 5 shows the result of another area of the blood sample, where two red blood cells (biconcave disks, upper left and lower right) and two white blood cells (upper right and lower left) are covered by the scanning range. Other parameters remain the same as in Fig. 4. Both cell types are well reconstructed and readily distinguishable.

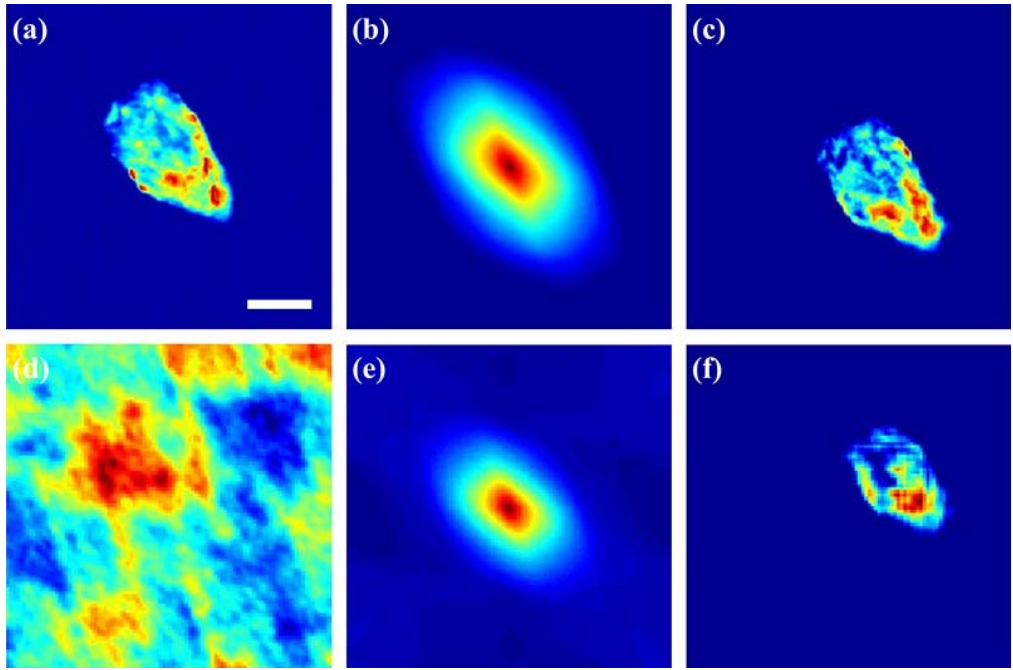

Fig. 4 (a) Wide field fluorescence image of a white blood cell; (b) calculated autocorrelation of (a); (c) reconstructed object from the calculated autocorrelation (b); (d) typical two-dimensional fluorescence intensity map; (e) average autocorrelation of 30 intensity maps; (f) reconstructed image from the average autocorrelation (e). Scale bar: 10 μm.

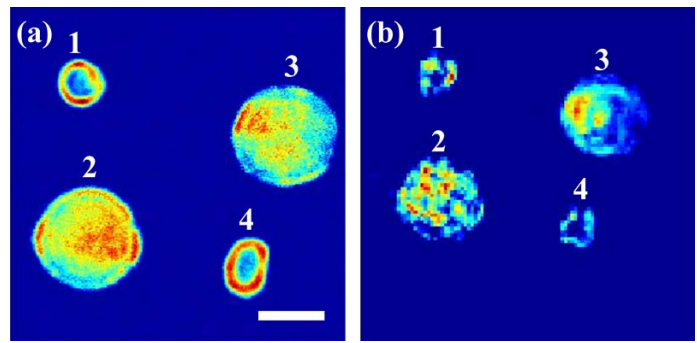

Fig. 5 (a) Wide field fluorescence image of two red blood cells and two white blood cells; (b) reconstructed image of the blood cells through chicken skin; scale bar: 10 μm

This result was obtained in transmission rather than in epi as in Ref [1]. On the epi side, the chicken skin tissue was illuminated by unscrambled laser beam, and generated autofluorescence comparable with the signal intensity from the blood cells; whereas in transmission side, the illuminating beam was diffused and broadened by the time it reached the second scattering layer, and the resulting autofluorescence from the tissue was negligible compared with the signal from the blood cells. Based on the fluorescence intensity measured with the EMCCD, the signal from the stained cells was 19 times stronger than the autofluorescence background of the scattering tissue layers in transmission configuration whereas the ratio is ~1 in the epi side.

## 4. Conclusion

We analyzed the effects on image quality of some implementation parameters of SSM, predominantly the distance between the scattering medium and the object. We showed that for the best performance of SSM, the scanning range of the correlated speckle pattern should be bigger than the dimension of the fluorescent object, and no other fluorescence sources should exist in close proximity outside the speckle scanning range. With these requirements satisfied, we obtained images of blood cells surrounded by chicken skin tissue. Cell type and shape are clearly identifiable with decent fidelity and subcellular resolution.

**Acknowledgement**

The authors thank Mr. Thomas Jean Victor Marie Lanvin for the preparation of chicken skin tissue. Xin Yang thanks Bertarelli foundation for partial financial support.

## Method

500 µl of blood was extracted from the ring finger of the experimenter, and kept in an EDTA tube to prevent clotting. The tube was centrifuged at 2000 r/min for 10 min, and the supernant was discarded. The rest of the blood was diluted with PBS at a ratio of 1:1, and centrifuged once more at 2000 r/min for 10 min. The supernant was discarded again, and the buffy coat was harvested. The harvested blood contained a mixture of leukocytes (white blood cells), thrombocytes (platelets) and some remaining erythrocytes (red blood cells). A blood smear was prepared following the coverslip technique [25], fixed with methanol, stained with Eosin-Y, and put between scattering media 1 and 2 as the imaging object.